\documentclass{elsart_sf}

\usepackage{epsfig}
\usepackage{amssymb, amsmath}
\RequirePackage{lineno}

\begin{document}
\begin{frontmatter}



\begin{center}
\title{Search for Neutrino Emission from Gamma-Ray Flaring Blazars with the ANTARES Telescope}
\end{center}
\author[UPV]{S.~Adri\'an-Mart\'inez},
\author[CPPM]{I. Al Samarai},
\author[Colmar]{A. Albert},
\author[UPC]{M.~Andr\'e},
\author[Genova]{M. Anghinolfi},
\author[Erlangen]{G. Anton},
\author[IRFU/SEDI]{S. Anvar},
\author[UPV]{M. Ardid},
\author[NIKHEF]{T.~Astraatmadja\thanksref{tag:1}},
\author[CPPM]{J-J. Aubert},
% ...

\newpage
\nopagebreak[3]
\address[UPV]{\scriptsize{Institut d'Investigaci\'o per a la Gesti\'o Integrada de les Zones Costaneres (IGIC) - Universitat Polit\`ecnica de Val\`encia. C/  Paranimf 1 , 46730 Gandia, Spain.}}\vspace*{0.15cm}
\nopagebreak[3]
\vspace*{-0.20\baselineskip}
\nopagebreak[3]
\address[CPPM]{\scriptsize{CPPM, Aix-Marseille Universit\'e, CNRS/IN2P3, Marseille, France}}\vspace*{0.15cm}
\nopagebreak[3]
\vspace*{-0.20\baselineskip}
\nopagebreak[3]
\address[Colmar]{\scriptsize{GRPHE - Institut universitaire de technologie de Colmar, 34 rue du Grillenbreit BP 50568 - 68008 Colmar, France }}\vspace*{0.15cm}
\nopagebreak[3]
\vspace*{-0.20\baselineskip}
\nopagebreak[3]
\address[UPC]{\scriptsize{Technical University of Catalonia, Laboratory of Applied Bioacoustics, Rambla Exposici\'o,08800 Vilanova i la Geltr\'u,Barcelona, Spain}}\vspace*{0.15cm}
\nopagebreak[3]
\vspace*{-0.20\baselineskip}
\nopagebreak[3]
\address[Genova]{\scriptsize{INFN - Sezione di Genova, Via Dodecaneso 33, 16146 Genova, Italy}}\vspace*{0.15cm}
\nopagebreak[3]
\vspace*{-0.20\baselineskip}
\nopagebreak[3]
\address[Erlangen]{\scriptsize{Friedrich-Alexander-Universit\"at Erlangen-N\"urnberg, Erlangen Centre for Astroparticle Physics, Erwin-Rommel-Str. 1, 91058 Erlangen, Germany}}\vspace*{0.15cm}
\nopagebreak[3]
\vspace*{-0.20\baselineskip}
\nopagebreak[3]
\address[IRFU/SEDI]{\scriptsize{Direction des Sciences de la Mati\`ere - Institut de recherche sur les lois fondamentales de l'Univers - Service d'Electronique des D\'etecteurs et d'Informatique, CEA Saclay, 91191 Gif-sur-Yvette Cedex, France}}\vspace*{0.15cm}
\nopagebreak[3]
\vspace*{-0.20\baselineskip}
\nopagebreak[3]
\address[NIKHEF]{\scriptsize{Nikhef, Science Park,  Amsterdam, The Netherlands}}\vspace*{0.15cm}
\nopagebreak[3]
\vspace*{-0.20\baselineskip}
\nopagebreak[3]
\address[APC]{\scriptsize{APC - Laboratoire AstroParticule et Cosmologie, UMR 7164 (CNRS, Universit\'e Paris 7 Diderot, CEA, Observatoire de Paris) 10, rue Alice Domon et L\'eonie Duquet 75205 Paris Cedex 13,  France}}\vspace*{0.15cm}
\nopagebreak[3]
\vspace*{-0.20\baselineskip}
\nopagebreak[3]
\address[LAM]{\scriptsize{LAM - Laboratoire d'Astrophysique de Marseille, P\^ole de l'\'Etoile Site de Ch\^ateau-Gombert, rue Fr\'ed\'eric Joliot-Curie 38,  13388 Marseille Cedex 13, France }}\vspace*{0.15cm}
\nopagebreak[3]
\vspace*{-0.20\baselineskip}
\nopagebreak[3]
\address[Bologna]{\scriptsize{INFN - Sezione di Bologna, Viale Berti-Pichat 6/2, 40127 Bologna, Italy}}\vspace*{0.15cm}
\nopagebreak[3]
\vspace*{-0.20\baselineskip}
\nopagebreak[3]
\address[Bologna-UNI]{\scriptsize{Dipartimento di Fisica dell'Universit\`a, Viale Berti Pichat 6/2, 40127 Bologna, Italy}}\vspace*{0.15cm}
\nopagebreak[3]
\vspace*{-0.20\baselineskip}
\nopagebreak[3]
\address[Pisa]{\scriptsize{INFN - Sezione di Pisa, Largo B. Pontecorvo 3, 56127 Pisa, Italy}}\vspace*{0.15cm}
\nopagebreak[3]
\vspace*{-0.20\baselineskip}
\nopagebreak[3]
\address[IFIC]{\scriptsize{IFIC - Instituto de F\'isica Corpuscular, Edificios Investigaci\'on de Paterna, CSIC - Universitat de Val\`encia, Apdo. de Correos 22085, 46071 Valencia, Spain}}\vspace*{0.15cm}
\nopagebreak[3]
\vspace*{-0.20\baselineskip}
\nopagebreak[3]
\address[Roma]{\scriptsize{INFN -Sezione di Roma, P.le Aldo Moro 2, 00185 Roma, Italy}}\vspace*{0.15cm}
\nopagebreak[3]
\vspace*{-0.20\baselineskip}
\nopagebreak[3]
\address[Roma-UNI]{\scriptsize{Dipartimento di Fisica dell'Universit\`a La Sapienza, P.le Aldo Moro 2, 00185 Roma, Italy}}\vspace*{0.15cm}
\nopagebreak[3]
\vspace*{-0.20\baselineskip}
\nopagebreak[3]
\address[Clermont-Ferrand]{\scriptsize{Clermont Universit\'e, Universit\'e Blaise Pascal, CNRS/IN2P3, Laboratoire de Physique Corpusculaire, BP 10448, 63000 Clermont-Ferrand, France}}\vspace*{0.15cm}
\nopagebreak[3]
\vspace*{-0.20\baselineskip}
\nopagebreak[3]
\address[GEOAZUR]{\scriptsize{G\'eoazur - Universit\'e de Nice Sophia-Antipolis, CNRS/INSU, IRD, Observatoire de la C\^ote d'Azur and Universit\'e Pierre et Marie Curie, BP 48, 06235 Villefranche-sur-mer, France}}\vspace*{0.15cm}
\nopagebreak[3]
\vspace*{-0.20\baselineskip}
\nopagebreak[3]
\address[Bari]{\scriptsize{INFN - Sezione di Bari, Via E. Orabona 4, 70126 Bari, Italy}}\vspace*{0.15cm}
\nopagebreak[3]
\vspace*{-0.20\baselineskip}
\nopagebreak[3]
\address[COM]{\scriptsize{COM - Centre d'Oc\'eanologie de Marseille, CNRS/INSU et Universit\'e de la M\'editerran\'ee, 163 Avenue de Luminy, Case 901, 13288 Marseille Cedex 9, France}}\vspace*{0.15cm}
\nopagebreak[3]
\vspace*{-0.20\baselineskip}
\nopagebreak[3]
\address[LNS]{\scriptsize{INFN - Laboratori Nazionali del Sud (LNS), Via S. Sofia 62, 95123 Catania, Italy}}\vspace*{0.15cm}
\nopagebreak[3]
\vspace*{-0.20\baselineskip}
\nopagebreak[3]
\address[UPS]{\scriptsize{Univ Paris-Sud , 91405 Orsay Cedex, France}}\vspace*{0.15cm}
\nopagebreak[3]
\vspace*{-0.20\baselineskip}
\nopagebreak[3]
\address[KVI]{\scriptsize{Kernfysisch Versneller Instituut (KVI), University of Groningen, Zernikelaan 25, 9747 AA Groningen, The Netherlands}}\vspace*{0.15cm}
\nopagebreak[3]
\vspace*{-0.20\baselineskip}
\nopagebreak[3]
\address[IRFU/SPP]{\scriptsize{Direction des Sciences de la Mati\`ere - Institut de recherche sur les lois fondamentales de l'Univers - Service de Physique des Particules, CEA Saclay, 91191 Gif-sur-Yvette Cedex, France}}\vspace*{0.15cm}
\nopagebreak[3]
\vspace*{-0.20\baselineskip}
\nopagebreak[3]
\address[Pisa-UNI]{\scriptsize{Dipartimento di Fisica dell'Universit\`a, Largo B. Pontecorvo 3, 56127 Pisa, Italy}}\vspace*{0.15cm}
\nopagebreak[3]
\vspace*{-0.20\baselineskip}
\nopagebreak[3]
\address[NIOZ]{\scriptsize{Royal Netherlands Institute for Sea Research (NIOZ), Landsdiep 4,1797 SZ 't Horntje (Texel), The Netherlands}}\vspace*{0.15cm}
\nopagebreak[3]
\vspace*{-0.20\baselineskip}
\nopagebreak[3]
\address[Bamberg]{\scriptsize{Dr. Remeis-Sternwarte and ECAP, Universit\"at Erlangen-N\"urnberg,  Sternwartstr. 7, 96049 Bamberg, Germany}}\vspace*{0.15cm}
\nopagebreak[3]
\vspace*{-0.20\baselineskip}
\nopagebreak[3]
\address[UU]{\scriptsize{Universiteit Utrecht, Faculteit Betawetenschappen, Princetonplein 5, 3584 CC Utrecht, The Netherlands}}\vspace*{0.15cm}
\nopagebreak[3]
\vspace*{-0.20\baselineskip}
\nopagebreak[3]
\address[UvA]{\scriptsize{Universiteit van Amsterdam, Instituut voor Hoge-Energie Fysika, Science Park 105, 1098 XG Amsterdam, The Netherlands}}\vspace*{0.15cm}
\nopagebreak[3]
\vspace*{-0.20\baselineskip}
\nopagebreak[3]
\address[MSU]{\scriptsize{Moscow State University,Skobeltsyn Institute of Nuclear Physics,Leninskie gory, 119991 Moscow, Russia}}\vspace*{0.15cm}
\nopagebreak[3]
\vspace*{-0.20\baselineskip}
\nopagebreak[3]
\address[Catania]{\scriptsize{INFN - Sezione di Catania, Viale Andrea Doria 6, 95125 Catania, Italy}}\vspace*{0.15cm}
\nopagebreak[3]
\vspace*{-0.20\baselineskip}
\nopagebreak[3]
\address[Catania-UNI]{\scriptsize{Dipartimento di Fisica ed Astronomia dell'Universit\`a, Viale Andrea Doria 6, 95125 Catania, Italy}}\vspace*{0.15cm}
\nopagebreak[3]
\vspace*{-0.20\baselineskip}
\nopagebreak[3]
\address[WIN]{\scriptsize{University of Wisconsin - Madison, 53715, WI, USA}}\vspace*{0.15cm}
\nopagebreak[3]
\vspace*{-0.20\baselineskip}
\nopagebreak[3]
\address[ISS]{\scriptsize{Institute for Space Sciences, R-77125 Bucharest, M\u{a}gurele, Romania     }}\vspace*{0.15cm}
\nopagebreak[3]
\vspace*{-0.20\baselineskip}
\nopagebreak[3]
\address[IPHC]{\scriptsize{IPHC-Institut Pluridisciplinaire Hubert Curien - Universit\'e de Strasbourg et CNRS/IN2P3  23 rue du Loess, BP 28,  67037 Strasbourg Cedex 2, France}}\vspace*{0.15cm}
\nopagebreak[3]
\vspace*{-0.20\baselineskip}
\nopagebreak[3]
\address[ITEP]{\scriptsize{ITEP - Institute for Theoretical and Experimental Physics, B. Cheremushkinskaya 25, 117218 Moscow, Russia}}\vspace*{0.15cm}
\nopagebreak[3]
\vspace*{-0.20\baselineskip}
\nopagebreak[3]
\address[Genova-UNI]{\scriptsize{Dipartimento di Fisica dell'Universit\`a, Via Dodecaneso 33, 16146 Genova, Italy}}\vspace*{0.15cm}
\nopagebreak[3]
\vspace*{-0.20\baselineskip}

\begin{abstract}
The ANTARES telescope is well-suited to detect neutrinos produced in astrophysical
transient sources as it can observe a full hemisphere of the sky at all times with a high
duty cycle. Radio-loud active galactic nuclei with jets pointing almost directly towards
the observer, the so-called blazars, are particularly attractive potential neutrino point
sources. The all-sky monitor LAT on board the Fermi satellite probes the variability of any given 
gamma-ray bright blazar in the sky on time scales of hours to months. Assuming hadronic models, a strong correlation between the gamma-ray and the neutrino
fluxes is expected. Selecting a narrow time window on the assumed neutrino production
period can significantly reduce the background.

An unbinned method based on the minimization of a likelihood ratio was applied to a
subsample of data collected in 2008 (61 days live time). By searching for neutrinos
during the high state periods of the AGN light curve, the sensitivity to these sources
was improved by about a factor of two with respect to a standard time-integrated point
source search. First results on the search for neutrinos associated with ten bright and variable Fermi
sources are presented.
\end{abstract}

\end{frontmatter}
\linenumbers

\section{Introduction}
Neutrinos are unique messengers to study the high-energy universe as they are neutral and 
stable, interact weakly and therefore travel directly from their point of creation to the Earth without absorption. 
Neutrinos could play an important role in understanding the mechanisms of cosmic ray acceleration and their 
detection from a cosmic source would be a direct evidence of the presence of hadronic acceleration. 
The production of high-energy neutrinos has been proposed for several kinds
of astrophysical sources, such as active galactic nuclei (AGN), gamma-ray bursters (GRB), supernova
remnants and microquasars, in which the acceleration of hadrons may occur (see Ref.~\cite{bib:Becker} for a review). 

Flat-Spectrum Radio Quasars (FSRQs) and BL Lacs, classified as AGN blazars, exhibit relativistic jets pointing almost directly towards
the Earth and are some of the most violent variable high energy phenomena in the Universe~\cite{bib:Blazars}. These sources are among the most likely sources of the observed ultra high energy
cosmic rays. Blazars typically display spectra with enhanced emission over two 
energy ranges: the IR/X-ray and MeV/TeV peaks. The lower energy peak is generally agreed to be the product of synchrotron radiation from accelerated electrons. However, 
the origin of the higher energy peak remains to be clarified. In leptonic models~\cite{bib:AGNleptonic}, inverse Compton scattering of synchrotron 
photons (or other ambient photons) by accelerated electrons generates this high energy emission. In hadronic models~\cite{bib:AGNhadronic}, 
MeV-TeV gamma-rays and high energy neutrinos are produced through hadronic interactions of the high energy cosmic rays with radiation or gas clouds surrounding the source.  
In the latter scenario, a strong correlation between the gamma-ray and the neutrino fluxes is expected. The gamma-ray light curves of bright blazars measured by the 
LAT instrument on board the Fermi satellite reveal important time variability on timescales of hours to several weeks, with intensities
much larger than the typical flux of the source in its quiescent state~\cite{bib:FermiLATAGNvariability}. 

This paper presents the results of the first time-dependent search for cosmic neutrino sources by the ANTARES telescope. The data sample used in
this analysis and the comparison to Monte Carlo simulations are described in Section 2,
together with a discussion on the systematic uncertainties. The point source search
algorithm used in this time-dependent analysis is explained in Section 3. The search results are presented in
Section 4 for ten selected candidate sources.

\section{ANTARES}

The ANTARES Collaboration completed the construction of a neutrino
telescope in the Mediterranean Sea with the connection of its twelfth detector line
in May 2008~\cite{bib:Antares}. The telescope is located 40 km off the Southern coast of France
(42$^{\circ}$48'N, 6$^{\circ}$10'E) at a depth of 2475 m. It comprises a three-dimensional array of
photomultipliers housed in glass spheres (optical modules~\cite{bib:OM}), distributed along twelve
slender lines anchored at the sea bottom and kept taut by a buoy at the top. 
Each line is composed of 25 storeys of triplets of optical modules (OMs), each housing one 10-inch photomultiplier. 
The lines are subject to the sea currents and can change shape and orientation. A positioning system 
based on hydrophones, compasses and tiltmeters is used to monitor the detector geometry with an accuracy of $~10$~cm.

The main goal of the experiment is to search for high energy neutrinos with energies greater than 100~GeV by detecting muons produced 
by the neutrino charged current interaction in the vicinity 
of the detector. Due to the large background from downgoing atmospheric muons, the telescope is optimized 
for the detection of upgoing muons as only they can originate from neutrinos.

Muons induce the emission of Cherenkov light in the sea water. The arrival time and intensity of the Cherenkov light on the OMs are digitized into hits and transmitted to shore. 
Events containing muons are selected from the continuous deep sea optical backgrounds due to natural radioactivity and bioluminescence. A detailed description of the detector 
and the data acquisition is given in~\cite{bib:Antares,bib:antaresdaq}.

The arrival times of the hits are calibrated as described in~\cite{bib:TimeCalib}. A L1 hit is defined either as a high-charge hit, or as hits separated by less 
than 20~ns on OMs of the same storey. At least five L1 hits are required throughout the detector within a time window of 2.2~$\mu$s, 
with the relative photon arrival times being compatible with the light coming from a relativistic particle. Independently, 
events which have L1 hits on two sets of adjacent or next-to-adjacent floors are also selected. 

The data used in this analysis were taken in the period from September 6 to
December 31, 2008 (54720 to 54831 modified Julian days, MJD) with the twelve line detector. This period overlaps with the availability of the first data from the LAT instrument 
onboard the Fermi satellite. The corresponding effective live time is 60.8 days.
Atmospheric neutrinos are the main source of background in the search for astrophysical neutrinos. These upgoing neutrinos 
are produced by the interaction of cosmic rays in the Earth's atmosphere. To account for this background, neutrino events were simulated according to the parametrization of 
the atmospheric neutrino flux from Ref.~\cite{bib:horandel}. Only charged 
current interactions of muon neutrinos and antineutrinos were considered. An additional source of background 
is due to downgoing atmospheric muons mis-reconstructed as upgoing. Downgoing atmospheric muons were simulated with the MUPAGE package~\cite{bib:Mupage}. 
In both cases, the Cherenkov light was propagated taking into account light absorption and scattering in sea water~\cite{bib:light}.
 
From the timing and position information of the hits, muon tracks are reconstructed
using a multi-stage fitting procedure, based on Ref.~\cite{bib:AAfit}. The initial fitting stages
provide the hit selection and starting point for the final fit. The final stage consists of a
maximum likelihood fit of the observed hit times and includes the contribution of optical
background hits. 

\begin{figure}[ht!]
\centering
\includegraphics[width=0.9\textwidth]{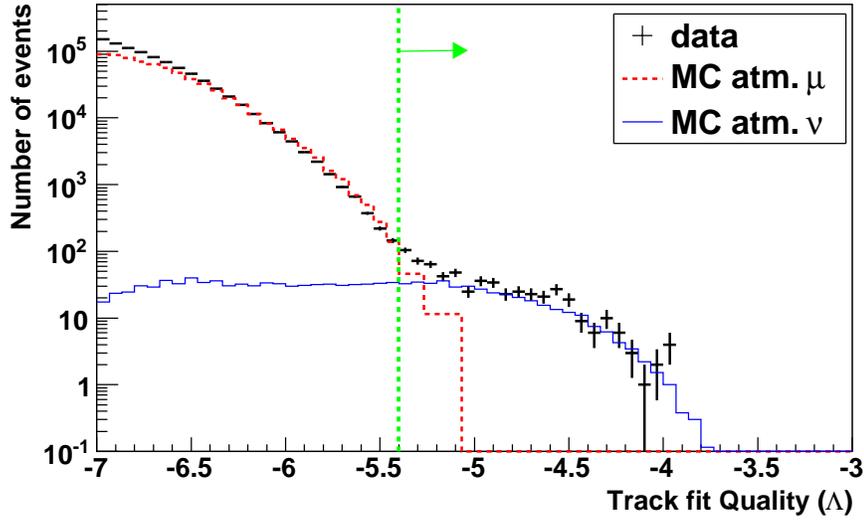}
\caption{Track fit quality ($\Lambda$) distribution for upgoing events
in data (dots) and Monte Carlo samples (atmospheric muons: dashed line; atmospheric neutrinos: continuous line). Events are selected with an error estimate lower 
than 1 degree. The green dashed vertical line corresponds to the optimized event selection ($\Lambda>-5.4$).}
\label{fig:FitQuality}
\end{figure}

Upgoing tracks are also required to have a good reconstruction quality. The latter is quantified by a parameter, $\Lambda$ which is based on the
value of the likelihood function obtained for the fitted muon (see Ref.~\cite{bib:AAfit} for details).
The cumulative distribution of $\Lambda$ for muons reconstructed as upgoing is shown
in Figure~\ref{fig:FitQuality} along with the simulated contributions from atmospheric muons and neutrinos.
The angular uncertainty obtained from the muon track fit is required to be smaller than 1 degree. 
For this analysis, events are selected with $\Lambda>-5.4$. This value results in an optimal compromise between the atmospheric neutrino and muon background reduction and 
the efficiency of the cosmic neutrino signal with an assumed spectrum proportional to $E_{\nu}^{-2}$, 
where $E_{\nu}$ is the neutrino energy, which gives the best 5$\sigma$ discovery potential. The resulting sample consists of 628 events obtained in 60.8 days. 
The simulations indicate that the selected sample contains 60~\% atmospheric neutrinos; the rest being mis-reconstructed atmospheric muons.

\begin{figure}[ht!]
\centering
\includegraphics[width=0.8\textwidth]{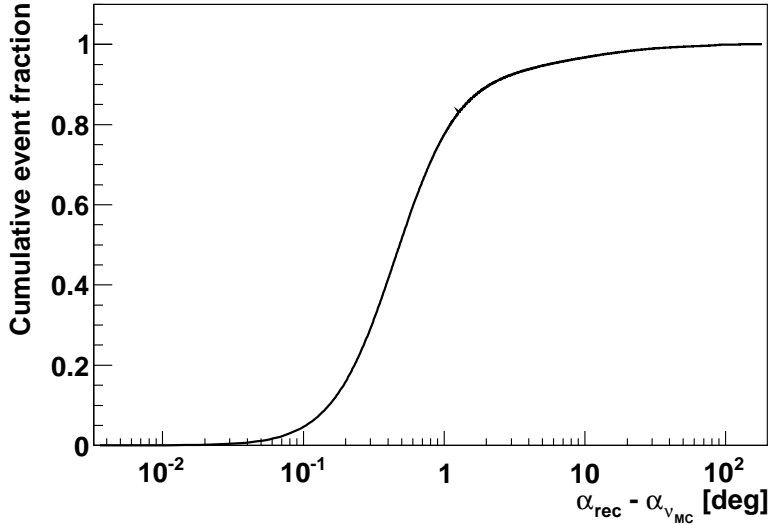}
\caption{Cumulative distribution of the angle between the true Monte Carlo neutrino direction ($\alpha_{\nu_{MC}}$) and the reconstructed
muon direction ($\alpha_{rec}$) for an E$_{\nu}^{-2}$ flux of upgoing neutrino events selected for this analysis.}
\label{fig:Angres}
\end{figure}

The angular resolution of the reconstructed neutrino direction can not be determined directly from the data and has to be estimated from simulation. However,
 comparison of data and Monte Carlo in which the time accuracy of the hits was degraded by up to 3~ns constrains the 
uncertainty of the angular resolution to about 0.1$^{\circ}$~\cite{bib:AAfitps}. Figure~\ref{fig:Angres} shows the cumulative distribution of the 
angular difference between the reconstructed muon direction and the neutrino direction for an assumed spectrum proportional to $E_{\nu}^{-2}$. For the 
considered period, the median resolution is estimated to be 0.5 $\pm$ 0.1 degrees.

\begin{figure}[ht!]
\centering
\includegraphics[width=0.8\textwidth]{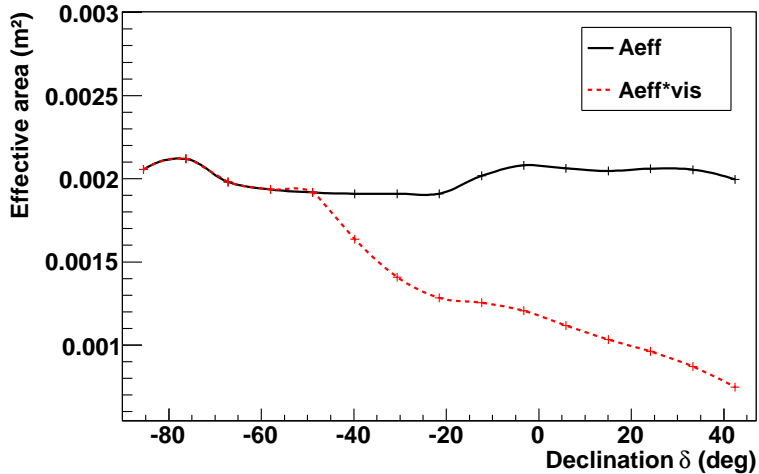}
\caption{ANTARES muon neutrino and antineutrino effective area (continuous line) as a function of the declination of the source computed from the Monte Carlo simulation for an E$_{\nu}^{-2}$ 
flux of upgoing muons selected for this analysis. The product of the effective area by the visibility (i.e. fraction of the time the source is visible at the ANTARES location) is shown with the dashed line.}
\label{fig:Seff}
\end{figure}

The effective area for muon neutrinos is defined as the ratio between the rate of selected neutrino events and the cosmic neutrino flux. Figure~\ref{fig:Seff} shows the muon 
neutrino and antineutrino effective area 
of the ANTARES telescope as a function of the declination of the source, after 
integrating over the energy with an assumed spectrum proportional to $E_{\nu}^{-2}$ between 10~GeV and 10~PeV. In the flux limits (see Section 4), a conservative uncertainty on the 
detection efficiency of about 30~\% was taken into account. This number includes contributions on the uncertainty of the sea water optical parameters~\cite{bib:light} and the OM 
properties such as efficiency and angular acceptance.

\section{Time-Dependent Search Algorithm}
The time-dependent point source analysis is performed using an unbinned method based on a
likelihood ratio maximization. The data are parametrized as a
mixture of signal and background. The goal is to determine, at a given point in the sky and at a given
time, the relative contribution of each component and to calculate the probability to have a signal
above background in a given model. The likelihood ratio, $\lambda$, is the logarithm of the ratio of the probability density for the hypothesis 
of signal and background ($H_{sig+bkg}$) over the probability density of only background ($H_{bkg}$):

\begin{equation}\label{eq:EQ_likelihood}
\lambda=\sum_{i=1}^{N} log\frac{P(x_{i}|H_{sig+bkg})}{P(x_{i}|H_{bkg})} = \sum_{i=1}^{N} log\frac{\frac{n_{sig}}{N}P_{sig}(\alpha_{i},t_{i}) + (1-\frac{n_{sig}}{N})P_{bkg}(\delta_{i},t_{i})}{P_{bkg}(\alpha_{i},t_{i})}
\end{equation}

where $n_{sig}$ is the unknown number of signal events determined by the fit and N is the total number of events in the considered data sample. 
$P_{sig}(\alpha_{i},t_{i})$ and  $P_{bkg}(\delta_{i},t_{i})$ are the probability density functions (PDF) for signal and background respectively. 
For a given event \textit{i}, $t_{i}$, $\delta_{i}$ and $\alpha_{i}$ represent the time of the event, its declination and the angular separation 
from the source under consideration. 

\begin{figure}[ht!]
\centering
\includegraphics[width=0.8\textwidth]{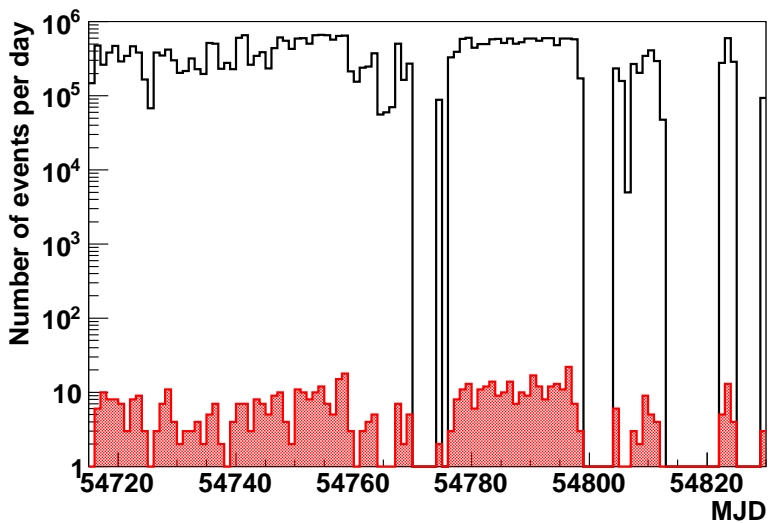}
\caption{Time distribution of the reconstructed events. Upper histogram (black line): distribution of all reconstructed events. Bottom filled histogram (red): distribution of selected upgoing events.
}
\label{fig:TimeDistri}
\end{figure}

The probability densities $P_{sig}$ and $P_{bkg}$ are factorized into a purely directional and a purely time-related component. 
The shape of the time PDF for the signal event is extracted directly from the gamma-ray light curve assuming 
proportionality between the gamma-ray and the neutrino fluxes. It is assumed that the muon neutrino velocity in vacuum is equal to that of light in vacuum. 
For signal events, the directional PDF is described by the one 
dimensional point spread function (PSF), which is the probability density of reconstructing an event at an angular distance $\alpha$ from the true source position. 
The directional and time PDF for the background are derived from the data using the observed declination distribution of the selected events 
and the observed one-day binned time distribution of all the reconstructed muons respectively. 
Figure~\ref{fig:TimeDistri} shows the time distribution of all the reconstructed events and the selected upgoing events for this analysis. Once normalized to an integral 
equal to 1, the distribution for all reconstructed events is used directly as the time PDF for the background.
Empty bins in the histograms correspond to periods with no data taking (i.e. detector in maintenance) or with very poor quality data (high bioluminescence or 
bad calibration).

The statistical interpretation of the search result relies on simulated pseudo experiments (PE) in which the background events are randomly generated by 
sampling the declination and the time from the parametrization $P_{bkg}(\delta_{i},t_{i})$ and the right ascension from a uniform distribution. Events from a 
neutrino point source are simulated 
by adding events around the desired coordinates according to the point spread function and the time distribution of the studied source. Systematic uncertainties 
(cf Section 2) are incorporated directly into the pseudo experiment generation. 

The null hypothesis corresponds to $n_{sig}=0$. The obtained value of $\lambda_{data}$ on the data is then 
compared to the distribution of $\lambda(n_{sig}=0)$. Large values of $\lambda_{data}$ compared to the distribution of $\lambda(n_{sig}=0)$ reject the null hypothesis 
with a confidence level (C.L.) equal to the fraction of the number of PE above $\lambda_{data}$. The fraction of 
PE for which $\lambda(n_{sig}=0)$ is above $\lambda_{data}$ is referred to as the p-value. The discovery potential is then defined as the average number of signal events required to 
achieve a p-value lower than 5$\sigma$ in 50~\% of the PEs. In the same way, the sensitivity is defined as the average signal required to obtain a p-value 
less than that of the median of the $\lambda(n_{sig}=0)$ distribution in 90~\% of the PEs. In the absence of evidence of a signal, an upper limit on the 
neutrino fluence is obtained and defined as the integral in energy and time of the flux upper limit with an assumed energy spectrum proportional to 
$E_{\nu}^{-2}$ from 10 GeV to 10 PeV. The limits are calculated according to the classical (frequentist) method for upper limits~\cite{bib:Neyman}.

\begin{figure}[ht!]
\centering
\includegraphics[width=0.8\textwidth]{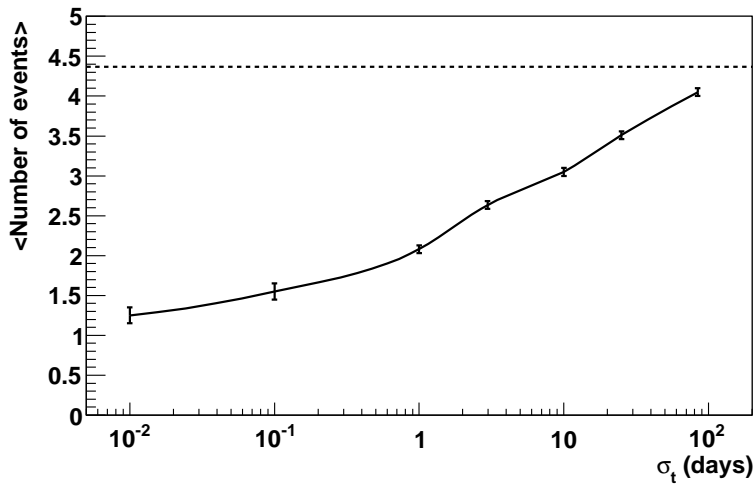}
\caption{Average number of events (solid line) required for a 5$\sigma$ discovery (50~\% probability) from a single source located at a
declination of -40$^{\circ}$ as a function of the width of the flare period ($\sigma_{t}$) for the 60.8 day analysis. These numbers are compared to that obtained without
using the timing information (dashed line).}
\label{fig:Nev5sigma}
\end{figure}

The performance of the time-dependent analysis was computed by applying this unbinned algorithm for a single source assuming a single square-shape flare with a width varying from 0.01 days to 84 days. 
The solid line in Figure~\ref{fig:Nev5sigma} shows the average number of events required for a discovery from one source located at a declination of -40$^{\circ}$ as 
a function of the width of the flare. The numbers in the black line are compared to that obtained without using the timing information (dashed line). The flare timing information yields 
an improvement of the discovery potential by about a factor 2-3 with respect to a standard time-integrated point source search~\cite{bib:AAfitps}.

\section{Search for Neutrino Emission from Gamma-Ray Flares}

The time-dependent analysis was applied to bright and variable Fermi blazar sources reported in the first-year Fermi LAT catalogue~\cite{bib:Fermicatalogue} and 
in the LBAS catalogue (LAT Bright AGN sample~\cite{bib:FermicatalogueAGN}). Sources were selected in the sky visible to ANTARES and that had at least one day binned gamma-ray flux
in the high state periods greater than 80x10$^{-8}$ photons cm$^{-2}$ s$^{-1}$ above 100~MeV and showed significant time variability on time scales of days to weeks in the 
studied time period. A source is assumed variable in the LBAS catalogue when the observation has a probability of less than 1~\% of being a steady source.
This list includes six flat-spectrum radio quasars and four BL-Lacs. Only four bright and nearby sources in the considered sample, PKS2155-304~\cite{bib:TeVsources1}, 
PKS1510-089~\cite{bib:TeVsources2}, 3C279~\cite{bib:TeVsources3} and WComae~\cite{bib:TeVsources4}, have been detected by the ground Cherenkov telescopes 
HESS, MAGIC or VERITAS. Table~\ref{tab:Sources} lists the characteristics of the ten selected sources.

\begin{table}[ht!]
\begin{center}
\begin{tabular}{|c|c|c|c|c|c|c|}
\hline
Name & {OFGL name} & Class & {RA [$^{o}$]} & {Dec [$^{o}$]} & Redshift \\
\hline
\hline
{PKS0208-512} & {J0210.8-5100} & FSRQ & 32.70 & -51.2 & 1.003 \\
\hline
{AO0235+164} & {J0238.6+1636} & BLLac & 39.65 & 16.61 & 0.940 \\
\hline
{PKS0454-234} & {J0457.1-2325} & FSRQ & 74.28 & -23.43 & 1.003 \\
\hline
{OJ287} & {J0855.4+2009} & BLLac & 133.85 & 20.09 & 0.306 \\
\hline
{WComae} & {J1221.7+28.14} & BLLAc & 185.43 & 28.14 & 0.102 \\
\hline
{3C273} & {J1229.1+0202} & FSRQ & 187.28 & 2.05 & 0.158 \\
\hline
{3C279} & {J1256.1-0548} & FSRQ & 194.03 & -5.8 & 0.536 \\
\hline
{PKS1510-089} & {J1512.7-0905} & FSRQ & 228.18 & -9.09 & 0.36 \\
\hline
{3C454.3} & {J2254.0+1609} & FSRQ & 343.50 & 16.15 & 0.859 \\
\hline
{PKS2155-304} & {J2158.8-3014} & BLLac & 329.70 & -30.24 & 0.116 \\
\hline
\end{tabular}
\caption{List of bright variable Fermi blazars selected for this analysis~\cite{bib:FermicatalogueAGN}.}
\label{tab:Sources}
\end{center}
\end{table}

\begin{figure}[ht!]
\centering
\includegraphics[width=0.9\textwidth]{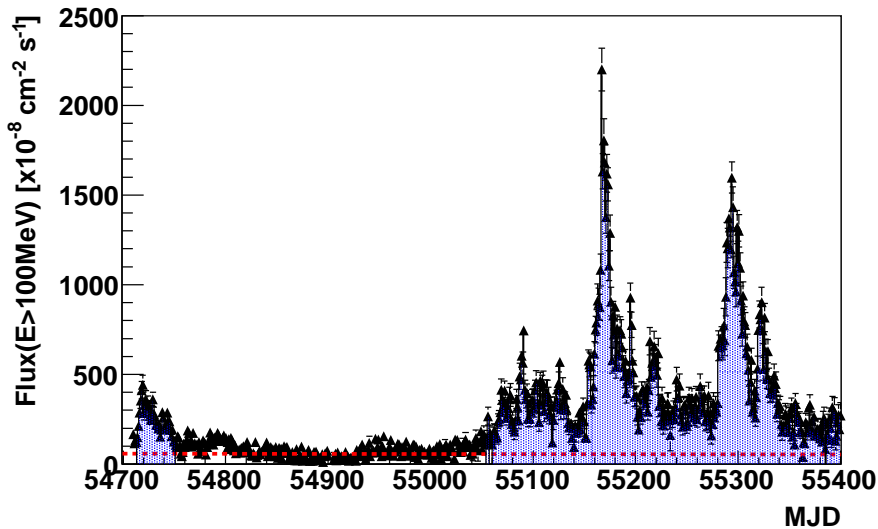}
\caption{Gamma-ray light curve (black points) of the blazar 3C454.3 measured by the LAT instrument onboard the Fermi satellite above 100~MeV for almost two years 
of data. The shaded histogram (blue) indicates the high state periods. The dashed line (red) represents the fitted baseline. 
}
\label{fig:3C454}
\end{figure}

The light curves published on the Fermi web page for the monitored sources~\cite{bib:Fermimonitored} are used for this analysis. They correspond to 
the one-day binned time evolution of the average gamma-ray flux above a threshold of 100~MeV since August 2008. The high state periods are defined 
using a simple and robust method based on three main steps. Firstly, the baseline is determined with an iterative linear fit. After each fit, bins more than two sigma ($\sigma_{BL}$) 
above the baseline (BL) are removed. Secondly, seeds for the high state periods are identified by searching for bins significantly above the baseline according to the criteria: 

\begin{equation}\label{eq:EQ_selection}
(F - \sigma_{F}) > (BL + 2*\sigma_{BL})~~and~~ F > (BL + 3*\sigma_{BL})
\end{equation}

where F and $\sigma_{F}$ represent the flux and the uncertainty on this flux for each bin, respectively. For each seed, the adjacent bins for which the emission is 
compatible with the flare are added if they satisfy: $(F - \sigma_{F}) > (BL + \sigma_{BL})$. Finally, an additional delay of 0.5 days is added before and after the flare in order to take into 
account that the precise time of the flare is not known (1-day binned light curve). With this definition, a flare has a width of at least two days. Figure~\ref{fig:3C454} shows 
the time distribution of the Fermi LAT gamma-ray light curve of 3C454.3 for almost two years of data and the corresponding selected high state periods. With the hypothesis that the neutrino 
emission follows the gamma-ray emission, the signal time PDF is simply the normalized light curve of only the high state periods. The third column of Table~\ref{tab:Results} 
lists the flaring periods for the ten sources found from September to December 2008.

\begin{figure}[ht!]
\centering
\includegraphics[width=0.9\textwidth]{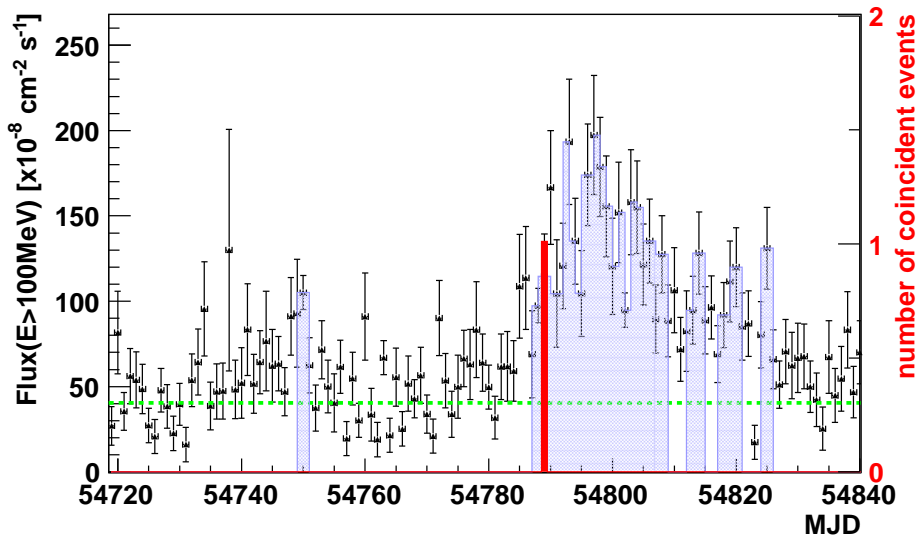}
\caption{Gamma-ray light curve (dots) of the blazar 3C279 measured by the LAT instrument onboard the Fermi satellite above 100 MeV. The light shaded histogram (blue)
indicates the high state periods. The dashed line (green) corresponds to the fitted baseline. The red histogram displays the time of the associated ANTARES neutrino event.
}
\label{fig:Result_3C279}
\end{figure}

The results of the search for coincidences between flares and neutrinos are listed in Table~\ref{tab:Results}. For nine sources, no coincidences are found. 
For 3C279, a single high-energy neutrino event is found in coincidence during a large flare in November 2008. Figure~\ref{fig:Result_3C279} shows the time 
distribution of the Fermi gamma-ray light curve of 3C279 and the time of the coincident neutrino event. This event was reconstructed with 89 hits distributed on 
ten lines with a track fit quality $\Lambda=-4.4$. The particle track direction is reconstructed at 0.56$^{\circ}$ from the source location. 
The pre-trial p-value is 1.0~\%. However, the post-trial probability computed taking into account the ten searches is 10~\%; this occurrence is thus compatible with 
a background fluctuation. In the absence of a discovery, upper limits on the neutrino fluence were computed and are shown in the last column of Table~\ref{tab:Results}.

\begin{table}[ht!]
      \begin{center}
      \begin{tabular}{|c|c|c|c|c|c|c|}
      \hline
            Source & Vis & {timePDF(MJD-54000)} & {LT} & {N($5\sigma$)} & $N_{obs}$ & {Fluence U.L.} \\
	    \hline
	    \hline
            {PKS0208-512} & 1.0 & {712-5,722-4,745-7,}& 8.8 & 4.5 & 0 & 2.8 \\
	                  &     &    {750-2,753-7,764-74,} &  &  &  &  \\
		          &     &    {820-2} &  &  &  &  \\
            \hline
	    {AO0235+164} & 0.41 & {710-33,738-43,746-64,} & 24.5 & 4.3 & 0 & 18.7\\
	    		&       & {766-74,785-7,805-8,}  &  &  &  & \\
				&       & {	810-2}  &  &  &  & \\
            \hline
	    {PKS1510-089} & 0.55 & {716-9,720-5,726-35,} & 4.9 & 3.8 & 0 & 2.8\\
	                  &       & {788-90,801-3}  &  &  &  & \\	    
            \hline
	    {3C273} & 0.49 & {714-6,716-8,742-5} & 2.4 & 2.5 & 0 & 1.1 \\
            \hline
	    {3C279} & 0.53 & {749-51,787-809,}& 13.8 & 5.0 & 1 & 8.2 \\
	                 &       & {812-5,817-21,824-6}  &  &  &  & \\
            \hline
	    {3C454.3} & 0.41 & {713-51,761-5,767-9,} & 30.8 & 4.4 & 0 & 23.5 \\
	                 &       & {784-801} &  &  &  & \\
            \hline
	    {OJ287} & 0.39 & {733-5,752-4,760-2,}& 4.3 & 3.9 & 0 & 3.4 \\
	                 &       & {768-70,774-6,800-2,}   &  &  &  & \\
		         &       & {814-6}   &  &  &  & \\
            \hline
	    {PKS0454-234} & 0.63 & {743-5,792-6,811-3} & 6.0 & 3.3 & 0 & 2.9 \\
            \hline
	    {WComae} & 0.33 & {726-9,771-3,790-2,}& 3.9 & 3.8 & 0 & 3.6 \\
	                &       & {795-7,815-7}    &  &  &  & \\
            \hline
	    {PKS2155-304} & 0.68 & {753-5,766-8,799-801,}& 3.1 & 3.7 & 0 & 1.6 \\
	                &       & {828-30}     &  &  &  & \\
      \hline
      \end{tabular}
      \caption{Results of the search for neutrino emission in the ten selected sources. The meaning of the columns is the following: Vis: fraction of the time the source is visible 
      at the ANTARES location; 
      timePDF: high state periods of the light curve; LT: corresponding ANTARES live time in days; N($5\sigma$): averaged number of events required for a 5$\sigma$ 
      discovery (50~\% probability); $N_{obs}$: number of observed events in time/angle coincidence with the gamma-ray emission. Fluence U.L.: Upper limit (90~\% C.L.) on 
      the neutrino fluence in GeV~cm$^{-2}$.}
      \label{tab:Results}
      \end{center}     
\end{table}

\section{Summary}
This paper presents the first time-dependent search for cosmic neutrinos using the data taken with the full twelve line ANTARES detector during the last four months of 2008. 
For variable sources, time-dependent point searches are much more sensitive than time-integrated searches due to the large reduction of the background. This 
search was applied to ten very bright and variable Fermi LAT blazars. One neutrino event was detected in time/direction coincidence with the gamma-ray emission in only one case, for a flare of 3C279
in November 2008, with a post-trial probability of 10~\%. Upper limits were obtained on the neutrino fluence for the ten selected sources.

\section{Acknowledgments}
The authors acknowledge the financial support of the funding agencies: 
Centre National de la Recherche Scientifique (CNRS), Commissariat 
\`a l'\'energie atomique et aux \'energies alternatives  (CEA), Agence
National de la Recherche (ANR), Commission Europ\'eenne (FEDER fund 
and Marie Curie Program), R\'egion Alsace (contrat CPER), R\'egion 
Provence-Alpes-C\^ote d'Azur, D\'e\-par\-tement du Var and Ville de 
La Seyne-sur-Mer, France; Bundesministerium f\"ur Bildung und Forschung 
(BMBF), Germany; Istituto Nazionale di Fisica Nucleare (INFN), Italy; 
Stichting voor Fundamenteel Onderzoek der Materie (FOM), Nederlandse 
organisatie voor Wetenschappelijk Onderzoek (NWO), the Netherlands; 
Council of the President of the Russian Federation for young scientists 
and leading scientific schools supporting grants, Russia; National 
Authority for Scientific Research (ANCS), Romania; Ministerio de Ciencia 
e Innovaci\'on (MICINN), Prometeo of Generalitat Valenciana and MultiDark, 
Spain. We also acknowledge the technical support of Ifremer, AIM and 
Foselev Marine for the sea operation and the CC-IN2P3 for the computing facilities.


\end{document}